\documentclass[aps,preprint,prc,superscriptaddress,showpacs,nofootinbib,floatfix,amssymb,amsfonts,amsmath]{revtex4-1}

\usepackage{graphicx}
\usepackage{epsf}
\usepackage{amsmath,amssymb}
\usepackage{bm}
\usepackage{color}
\usepackage{ulem}

\begin{document}

\title{A Dynamical Study of Fusion Hindrance with Nakajima-Zwanzig Projection Method}

\author{Yasuhisa Abe}
\affiliation{Research Center for Nuclear Physics(RCNP), Osaka University, 10-1 Mihogaoka, Ibaraki. 
567-0047 Osaka, Japan \email{abey@rcnp.osaka-u.ac.jp}}
\email{abey@rcnp.osaka-u.ac.jp}

\author{David Boilley}
\author{Quentin Hourdill\'e}
\affiliation{GANIL, CEA/DRF-CNRS/IN2P3, BP 55027, F-14076 Caen cedex 5, France}
\affiliation{Normandie Univ. Unicaen, Caen, France}

\author{Caiwan Shen}
\affiliation{School of Science, Huzhou University. Huzhou, 313000 Zhejiang, People's Republic of China}

\begin{abstract}
A new framework is proposed for the study of collisions between very heavy ions which lead 
to the synthesis of Super-Heavy Elements (SHE), to address the fusion hindrance phenomenon. 
The dynamics of the reaction is studied in terms of collective degrees of freedom undergoing 
relaxation processes with different time scales.  The Nakajima-Zwanzig projection operator 
method is employed to eliminate fast variable and derive a dynamical equation for the reduced 
system with only slow variables. There, the time evolution operator is renormalised and an 
inhomogeneous term appears, which represents a propagation of the given initial distribution. 
The term results in a slip to the initial values of the slow variables. We expect that gives a 
dynamical origin of parameter ``injection point $s$" introduced by Swiatecki et al in order to 
reproduce absolute values of measured cross sections for SHE. Formula for the slip is given in 
terms of physical parameters of the system, which confirms the results recently obtained with 
a Langevin equation. 
\end{abstract}

\maketitle

\section{Introduction}

Finding the limit of existence of nuclei is one of the challenging research programs in nuclear 
physics. It gives the limit of chemical elements, with which all the matter of the world is 
made.  Within the Liquid Drop Model (LDM), the limit of the atomic charge $Z$ is around 100, 
where the fission barrier becomes negligibly small. Beyond this limit, nuclei only owe their 
existence to the extra-stability provided by the shell correction energy due to quantum-mechanical 
structure effects in finite many-body Fermion systems.  They are called superheavy elements (SHE). 
Their quests in nature turned out unsuccessful, and then the prediction of their existence is to 
be justified by synthesis with nuclear reactions, but the attempts face a challenge due to the 
extremely low cross sections, at the order of the picobarn or even smaller for the heaviest 
elements produced by fusion-evaporation reactions \cite{Hof2009}. The reasons are twofold: First, 
fragility of the compound nuclei (CN) formed, that are supported by a very low fission barrier, 
and thus, undergo fission before cooling down through neutron evaporation.  Second, the fusion 
process is hindered with respect to models developed for the fusion of light nuclei. Fusion 
hindrance is observed in systems with $Z_1 \cdot Z_2$ being larger than $1\,600$ \cite{Sch1991}.
There is no commonly accepted explanation of the origin of the latter, while the former is 
quantitatively well described by the conventional statistical theory of decay \cite{Kewpie2}, 
with some ambiguities in physical parameters such as nuclear masses \cite{Lu2016}.     

An unveiling of the origin of the hindrance and its correct description is necessary to improve 
the predictive power of models describing the whole reaction leading to the synthesis of SHE. 
In all models, the fusion process consists in a sequence of two steps, i.e., overcoming of the 
Coulomb barrier and then formation of the CN, starting from the di-nucleus configuration of the 
hard contact of projectile and target nuclei. The necessity for the latter process is due to the 
fact that the contact configuration in heavy systems locates outside of the conditional saddle 
point, and thus, the system has to overcome the saddle point, after overcoming the Coulomb barrier \cite{Royer1985,Shen2002,Abe2002a,Abe2002b,Abe2003,FBD1,FBD2,Zagrebaev2005}.  In other words, 
the fusion probability is given by the product of the contact and the formation probabilities. 
The former factor is often called the capture probability or cross-section, while the latter 
is called formation factor or hindrance factor as in the Fusion-by-Diffusion model (FBD) \cite{FBD1,FBD2}.   

The dynamics of the latter process cannot be simply described by Newtonian mechanics, or its 
quantum version. For, at the hard contact of two nuclei, the di-nucleus system is internally 
excited. Most of the kinetic energy carried in by the incident channel is supposed to be already 
transformed into those of nucleonic motions, i.e., to be dissipated into heat. Therefore, the 
collective motions of the system are to be described by dissipation-fluctuation dynamics like 
in fission decay \cite{Kramers1940,Grange1983,Abe1986,Boilley1993,Abe1996,Fro1998}.  

However, first cross sections calculated with the Langevin model \cite{Aritomo1999,Bouriquet2004,Shen2008} 
as well as the Fusion-by-Diffusion model \cite{FBD1,FBD2}, turned out to be still too large, 
by a few orders of magnitude, compared with the measured cross sections in so-called cold fusion 
path for SHE. In order to reproduce the absolute value of the measured cross sections, an arbitrary 
``injection point parameter $s$'' was introduced in the latter model, shifting the contact configuration 
outward by about 2.0 fm. Such a shift makes the hindrance factor stronger and thus, the cross section 
even smaller by a few order of magnitude as desired \cite{FBD1,FBD2}. A systematics over several 
systems exhibited a regular behaviour \cite{Cap2011,Siwek2013}. A similar analysis was conducted 
on the so-called hot fusion path with an injection point parameter found to be incident-channel 
dependent or incident-energy dependent for systematic reproduction of the experiments 
\cite{Cap2012,Siwek2012,Hagino2018}. Those phenomenological results suggest that something 
is missing in theories of fusion mechanism hitherto developed.    

First attempts to explain the shift of the injection parameter to dynamical process from the 
di-nucleus to a mono-nucleus shape were related to the fast disappearance of the neck degree 
of freedom that affects the injection point and gives rise to an additional hindrance 
\cite{Abe2008,Abe2009,Boilley2011a,Boilley2011b,Liang2012,Zhu2013}. Later, combining of the 
elimination of both neck degree of freedom and momentum ones in case of strong friction lead 
to the incident-energy dependence of the initial slip, thus explaining the behaviour of the 
injection point parameter, systematically both for the cold and the hot fusion paths 
\cite{Boilley2019}. This is encouraging for theoretical predictions on on-going and/or 
future experiments for synthesis of heavier elements with various incident channels.

More generally, the elimination of fast variables in a multidimensional system to reduce 
the model to few slow variables is addressed in several seminal articles 
\cite{Chandrasekhar1943,Kampen1985} and textbooks \cite{Kampen1981,Risken1984,Gardiner1990}. 
It always leads to a slip of the initial condition of the slow variables 
\cite{Grad1963,McL1974,Sanctuary1977,Titulaer1978,Haake1983,Cox1995}. The purpose of the 
present article is to propose another framework for this problem.

In contrast with Ref. \cite{Boilley2019} which is based on the Langevin formalism 
\cite{Langevin1908}, we shall rather use here the so-called Fokker-Planck \cite{Fokker1914,Planck1917} 
formalism which is based on partial differential equations describing the time evolution of the 
probability density function in the phase space. To look at the problem from two viewpoints 
often provides deeper physical understanding of the dynamics. We, thus, expect that the 
present approach sheds another light on the problem how the fast neck degree affects the 
motion of the slow variables and gives rise to a new additional hindrance.

There are various Fokker-Planck type equations to study the diffusion of collective degrees of 
freedom dragged by potential forces. In this study, we shall consider  special forms describing 
the Brownian motion in an external field. When applied to position distributions, it is better 
known as Smoluchowski equation \cite{Smoluchowski1906}, and, when it also takes into account 
the conjugated momenta or velocity, it is called Klein-Kramers (K-K) equation \cite{Klein1921,Kramers1940}. 
The former is an approximation of the latter in the high viscosity limit. 

Applied to collective degrees of freedom describing the shape of the colliding nuclei from contact 
to compound shape, dragged by a macroscopic potential map based on the LDM, we aim to give a 
global base-line theory of fusion of heavy systems. The fate of the system is either fusion 
(CN formation) or to re-separation (quasi-fission decay).
Since the neck degree is distinctly (by factor several to one order of magnitude) faster 
than the other degrees of freedom \cite{Abe2009,Boilley2011a,Boilley2011b,Liang2012,Zhu2013}, 
the neck rapidly reaches the equilibrium point, or the equilibrium distribution at the very 
beginning of the dynamical processes. 

In order to rigorously eliminate the fast variable, we employ Nakajima-Zwanzig projection 
operator method \cite{Nakajima1958,Zwanzig1960} and derive a dynamical equation only for the 
slow variables and see how the initial slip arises. In  Section 2, the derivation of the reduced 
dynamical equation is given in the case of Smoluchowski equation with the fast (neck) and the 
slow (radial and mass-asymmetry) coordinates. For simplicity and facilitation of analytic 
calculations, we assume the potential (LDM energy surface) to be approximated by multi-dimensional 
parabola (Taylor expansion at the saddle point up to the second order) and the friction tensor 
for the  collective degrees to be coordinate-independent, i.e., be constant.  The coupling 
between the fast and the slow  variables is supposed to be weak.   Perturbative approximation with 
respect to the coupling enables us to obtain a reduced Fokker-Planck equation with simple analytic 
expressions for the evolution operator and the inhomogeneous term. The latter term is mostly neglected
in the scientific literature with a few exceptions 
\cite{Grad1963,McL1974,Sanctuary1977,Titulaer1978,Haake1983,Cox1995}, but turns out to play a 
key role in the present subject, giving rise to the $\it slip$ of the initial point and thus, 
an additional hindrance. In Section 3, we summarise the results, including those for K-K 
equation, which is equivalent to full Langevin equation.

\section{Elimination of fast variables by Nakajima-Zwanzig projection method and initial slip}

As mentioned in the introduction, the elimination of fast variables with the Nakajima-Zwanzig (N-Z) 
projection method \cite{Nakajima1958,Zwanzig1960} is general and can be applied to either K-K 
or Smoluchowski equation. Here, we shall only consider the Smoluchowski equation for general 
$N$-dimensional collective coordinates which describe shapes of the di-nucleus system. As an 
illustration, we eliminate the neck variable to derive a dynamical equation for the system with 
only slow variables and investigate possible effects of the eliminated variable to the motion of 
the slow ones. 
N-Z method is amenable to analytic calculations practically for cases that a coupling is separable 
with respect to fast and slow variables, like in Caldeira-Leggett model\cite{Leggett}.   In the present model, 
however, the coupling is in Smoluchowski type, as given below in 2.1.   Nevertheless, it turns out to be 
effectively bi-linear, helped by the quadratic potential, etc.   It, thus, is feasible to calculate 
the projection, i.e., the integration (trace) over the fast variable analytically, which results in a simple dynamical 
equation for the reduced system, as shown in detail in 2.2.

\subsection{Projectors}

The equation is 
\begin{equation}
\frac{\partial}{\partial t}w(\vec{q},t)=\mathcal{L}\cdot w(\vec{q},t),\label{Smol}
\end{equation}
with
\begin{equation}
\mathcal{L}  = \sum_{ij}\frac{\partial}{\partial q_{i}}\mu_{ij}\cdot\left( \frac{\partial V}
{\partial q_{j}}+T\frac{\partial}{\partial q_{j}}\right)  ,\label{L}
\end{equation}
where indexes $(i, j)$ denote all the fast and the slow variables, while indexes $(\alpha, \beta)$ 
used later only denote slow variables, i.e., the radial and mass-asymmetry coordinates. $V$ 
denotes LDM potential, while $\mu$ is the inverse of the friction tensor $\gamma$. $T$ is 
the temperature of the system. 

Smoluchowskian is rewritten as follows, with the premise of perturbative approximation with 
respect to the coupling between the fast and the slow coordinates,
\begin{equation}
\mathcal{L}  =\mathcal{L}_{0}+\mathcal{L}_{1},\quad     {\rm with} \quad  
\mathcal{L}_{0} =\mathcal{L}_{f}+\mathcal{L}_{s} \quad {\rm and} \quad \mathcal{L}_{1}
=\mathcal{L}_{fs}+\mathcal{L}_{sf},  \label{LSep}
\end{equation}
where 
\begin{eqnarray}
\mathcal{L}_{f} &=&\frac{\partial}{\partial q_{f}}\mu_{f}\left[\frac{\partial V}
{\partial q_{f}}+T\frac{\partial}{\partial q_{f}}\right],\\ 
\mathcal{L}_{s} &=& {\sum\limits_{\alpha\beta}}\frac{\partial}{\partial q_{\alpha}}
\mu_{\alpha\beta}\left[  \frac{\partial V}{\partial q_{\beta}}+T\frac{\partial}
{\partial q_{\beta}}\right],\\
\mathcal{L}_{fs} &=&{\sum\limits_{\alpha}}\frac{\partial}{\partial q_{f}} \mu_{f\alpha}
\left[  \frac{\partial V}{\partial q_{\alpha}}+T\frac{\partial}{\partial q_{\alpha}}\right], \\
\mathcal{L}_{sf} &=&{\sum\limits_{\alpha}}\frac{\partial}{\partial q_{\alpha}}\mu_{\alpha f}
\left[  \frac{\partial V}{\partial q_{f}}+T\frac{\partial}{\partial q_{f}}\right],
\end{eqnarray}
where $\mu_f$ represents the $\mu_{ff}$. For the sake of simplicity, we have only considered 
a single fast variable here. This could be easily generalised to any number of fast variables.

Assuming that the fast variable quickly converges towards an equilibrium distribution 
$\phi_0(q_f)$, we introduce the projection operators \cite{Nakajima1958,Zwanzig1960},
\begin{equation}
P  =\phi_{0}(q_{f})\cdot\int dq_{f},  \quad {\rm and} \quad Q  =1-P. \label{proj}
\end{equation}
Since $q_{f}$ becomes rapidly in equilibrium, non-equilibrium component disappears rapidly, 
and thus, life time of $Q$ space is short, compared with time scale of the slow variables.    

It is easy to show that projected distributions $Pw$ and $Qw$ satisfy
\begin{eqnarray}
\frac{\partial}{\partial t}Pw  &  =&P\mathcal{L}Pw+P\mathcal{L}Qw, \label{Pw}\\ 
\frac{\partial}{\partial t}Qw  &  =&Q\mathcal{L}Pw+Q\mathcal{L}Qw, \label{Qw}
\end{eqnarray}
which is rewritten into a closed equation for $Pw$,
\begin{equation}
\frac{\partial}{\partial t}Pw =  P\mathcal{L}Pw+P\mathcal{L}\cdot
e^{Q\mathcal{L}\cdot t}\int_{0}^{t}dt^{\prime}e^{-Q\mathcal{L}\cdot t^{\prime
}}Q\mathcal{L}Pw(t^{\prime}) +P\mathcal{L}\cdot e^{Q\mathcal{L}\cdot t}
\cdot Qw(\vec{q},0). \label{FullEq}
\end{equation}
So far, no approximation is made and therefore, this equation is equivalent to the 
original one, but is amenable to perturbative approximation. 

The last term of Eq. (\ref{FullEq}) carries the information of the $Q$ space component 
of the initial distribution function. Of course, it is natural to disregard it, because 
its effect quickly disappears as time goes. However, in systems with bifurcation or in 
open systems, it can dramatically change the fate of the system. Actually, in our case, 
the elimination results effectively in a $\it{slip}$ of the initial points of the slow 
variables, thus affecting the formation probability by a few order of magnitude, as is 
shown below. 

The second term of Eq. (\ref{FullEq})  exhibits a memory effect, which systematically 
appears when some variables are eliminated \cite{Chandrasekhar1943,Kampen1985,Kampen1981,Risken1984}. 
The method has been generally discussed in the field of statistical mechanics, but 
their interests are mostly to what dynamics the reduced system obeys.   We, however, 
are also interested in the inhomogeneous term in Eq. (\ref{FullEq}), which carries the 
$Q$ space component in the initial distribution.    

 \subsection{Renormalised Smoluchowski operator and initial slip}

To facilitate the calculations, we first diagonalise the coefficient matrix of the 
potential, although it is quite close to be diagonal in case of Two-Center-Parameterization (TCP) 
of di-nucleus system \cite{Maruhn1972,Sato1979}. Thus, the potential is as follows, 
\begin{equation}
V=\frac{1}{2}{\sum\limits_{ij}}c_{ij}q_{i}q_{j}\Rightarrow\frac{1}{2}
c_{f}q_{f}^{2}+\frac{1}{2}{\sum\limits_{\alpha}}c_{\alpha}q_{\alpha}^{2},
\label{2.2}
\end{equation}
then,
\begin{equation}
\frac{\partial V}{\partial q_{i}}=c_{i}q_{i}.  \label{2.3} 
\end{equation}
Note that $\mathcal{L}_{f}$ and $\mathcal{L}_{s}$ are operators solely in the fast and the 
slow variables, respectively, while  $\mathcal{L}_{1}$ is the coupling between them. 

In the following calculations, we use the basic properties of $\mathcal{L}_{f}$:
\begin{equation}
\int dq_{f}\mathcal{L}_{f}\cdot X  =0, \quad \forall X, \label{2.10} \\
\end{equation}
Naturally, distribution functions are supposed to be zero at the boundaries of $q_f$ variable. 
For the equilibrium distribution $\phi_0(q_f)$, we have that 
\begin{equation}
\mathcal{L}_{f}\phi_{0}(x) =0, \quad \text{with} \quad \phi_{0}(x)=\frac{1}
{\sqrt{2\pi}}e^{-\frac{1}{2}x^{2}}, \quad \text{and} \quad x=\sqrt{\frac{c_{f}}{T}}q_{f}
\end{equation}
The eigenvalues and eigenfunctions are  
\begin{equation}
\mathcal{L}_{f}\phi_{n}(x)  =-\lambda_{n}\phi_{n}(x), \quad \text{with} \quad \lambda_{n}
=n\mu_{f}c_{f} \quad \text{and} \quad \phi_{n}(x)=H_{n}(x)\cdot\phi_{0}(x).
\end{equation}
Here, $H_{n}(x)$ are the Hermite polynomials.

Hereafter, we calculate the terms in r.h.s. of Eq. (\ref{FullEq}) at the lowest order 
with respect to the coupling $\mathcal{L}_{1}$ between the fast and the slow variables.
For the first term, we simply have that 
\begin{equation}
P\mathcal{L}Pw  =\phi_{0}\mathcal{L}_{s}\cdot w(\vec{q}_{\alpha},t),\quad \text{with} 
\quad w(\vec{q}_{\alpha},t)  =\int dq_{f}w(\vec{q}_{i},t).
\end{equation}
This is just the Smoluchowski operator for the $(N-1)$ slow variables only.

The second term is already at the second order with respect to $\mathcal{L}_{1}$, due 
to the pre- and post-factors of the propagator. Therefore, we approximate the full Smoluchowski operator 
$\mathcal{L}$ in the propagator by $\mathcal{L}_{0}$, so that, $e^{Q\mathcal{L}\cdot t}
\simeq e^{Q\mathcal{L}_{0}\cdot t} $.
Noticing that $Q\mathcal{L}_{s}$ and $Q\mathcal{L}_{f}$ commute, $e^{Q(\mathcal{L}_{f}+\mathcal{L}_{s})
\cdot(t-t^{\prime})}=e^{Q\mathcal{L}_{f}\cdot(t-t^{\prime})}\cdot e^{Q\mathcal{L}_{s}\cdot(t-t^{\prime})}.$
Thus, with the properties of $\mathcal{L}_{f}$ given above, the memory kernel is governed
by the factor $e^{-\mu_{f}\cdot c_{f}\cdot (t-t^{\prime})}$. Thus, the duration of the 
memory effect is determined by the relaxation time of the fast neck degree. In the 
$t^{\prime}$ integration, we use Laplace approximation, taken into account that the 
integrand is slowly varying function,
\begin{equation}
e^{-\mu_{f}\cdot c_{f}\cdot (t-t^{\prime})}\simeq \delta(t^{\prime} -t)/(\mu_{f}\cdot c_{f}). 
\end{equation}
Then, the second term is rewritten as follows,
\begin{equation}
-\phi_{0}\frac{\partial}{\partial q_{\alpha}}\mu_{\alpha f}\frac{1}{\mu_{f}}\mu_{f\beta}
\cdot\left[c_{\beta}q_{\beta}+T\frac{\partial}{\partial q_{\beta}}\right]  w(\vec{q},t). 
\end{equation}

The first two terms of Eq. (\ref{FullEq}) are summed up into
\begin{equation}
\phi_{0}\frac{\partial}{\partial q_{\alpha}}\mu_{\alpha f}^{\text{eff}}\left[c_{\beta}
q_{\beta}+T\frac{\partial}{\partial q_{\beta}}\right]  w(\vec{q},t), \quad \text{with} 
\quad \mu_{\alpha\beta}^{\text{eff}}=\mu_{\alpha\beta}-\mu_{\alpha f}\frac{1}{\mu_{f}}
\mu_{f\beta.}. \label{mueff}
\end{equation}
This means that the elimination of the fast variable results in the renormalisation of 
Smoluchowski operator for the slow variables with $\mu_{\alpha\beta}^{\text{eff}}$:
\begin{equation}
\mathcal{L}^{\text{eff}}  =\frac{\partial}{\partial q_{\alpha}}\mu_{\alpha\beta}^{\text{eff}}
\left[c_{\beta}q_{\beta}+T\frac{\partial}{\partial q_{\beta}}\right].
\end{equation}

Now, we analyse the last term in Eq (\ref{FullEq}) which is most important in the present 
study. Using a perturbative expansion of the propagator $e^{Q\mathcal{L}\cdot t}
=e^{Q(\mathcal{L}_{0}+\mathcal{L}_{1})\cdot t}$ with respect to $\mathcal{L}_{1}$, the 
third term is given as follows,
\begin{equation}
P\mathcal{L}\cdot e^{Q\mathcal{L}\cdot t}Qw(0)=P(\mathcal{L}_{0} +\mathcal{L}_{1})
\cdot e^{Q\mathcal{L}_{0}\cdot t}\cdot\left[  1+\int_{0}^{t}dt^{\prime}e^{-Q\mathcal{L}_{0}\cdot t^{\prime}}Q\mathcal{L}_{1}e^{Q\mathcal{L}_{0}t^{\prime}}+\cdots\right]  \cdot Qw(0). 
\end{equation}
The zeroth order term is simply zero, and the first order term which does not vanish is,
\begin{eqnarray}
I_1 &= &P\mathcal{L}_{1}\cdot e^{Q\mathcal{L}_{0}\cdot t}Qw(0)  =P\mathcal{L}_{sf}
\cdot e^{Q\mathcal{L}_{0}\cdot t}Qw(0)\\
&=& \frac{\partial}{\partial q_{\alpha}}\mu_{\alpha f}\cdot c_{f}\phi_{0}\int dq_{f}
\text{ }q_{f}e^{Q\mathcal{L}_{s}\cdot t^{\prime}}\cdot e^{Q\mathcal{L}_{f}\cdot t}\cdot Qw(0).
\end{eqnarray}
Since $e^{Q\mathcal{L}_{f}\cdot t}=e^{\mathcal{L}_{f}\cdot t}$ , and $e^{Q\mathcal{L}_{s}
\cdot t}\cdot e^{Q\mathcal{L}_{f}\cdot t}\cdot Qw(0)=e^{\mathcal{L}_{s}\cdot t}\cdot 
e^{\mathcal{L}_{f}\cdot t}\cdot Qw(0)$,
\begin{eqnarray}
I_1 &=&\frac{\partial}{\partial q_{\alpha}} \mu_{\alpha f} \cdot c_{f} e^{\mathcal{L}_{s}
\cdot t}\phi_{0}\int dq_{f}\,q_{f}\left[e^{\mathcal{L}_{f}\cdot t}\phi(0)-\phi_{0}\right]
w(\vec{q},0)\\
 & = &\frac{\partial}{\partial q_{\alpha}} \mu_{\alpha f} c_{f} e^{\mathcal{L}_{s}\cdot t}
 \phi_{0}\int dq_f\, q_f \cdot \phi(q_{f},t) w(\vec{q}_{a},0),
\end{eqnarray}
where $w(q_{i},0)=w(q_{f},0) w(q_{\alpha},0)$ is assumed. And
\begin{equation}
\phi(q_{f},t)=e^{\mathcal{L}_{f}\cdot t}\cdot\phi(q_{f},t=0) 
\end{equation}
denotes the solution of 1-D Smoluchowski equation, which is well known 
\cite{Chandrasekhar1943,Kampen1985,Kampen1981,Risken1984}: 
\begin{equation}
\phi(q_{f},t)  =\frac{1}{\sqrt{2\pi\sigma(t)}}\exp\left[-\frac{(q_{f}-\left\langle q_{f}(t)
\right\rangle )^{2}}{2\sigma(t)}\right], 
\end{equation}
with
\begin{equation}
\left\langle q_{f}(t)\right\rangle=q_{f0}e^{-\mu_{f}c_{f}\cdot t} \quad \text{and} \quad 
\sigma(t)  =\sigma(0)e^{-2\mu_{f}c_{f}\cdot t}+\frac{T}{c_{f}}\left[1-e^{-2\mu_{f}c_{f}\cdot t}\right],
\end{equation}
where $q_{f0}$ is the initial value of the fast variable.
Finally, integrated over $q_{f}$, the term $ I_1 $ becomes
\begin{equation}
\vec{B}(t)^{T}\cdot\frac{\partial}{\partial\vec{q}}e^{\mathcal{L}_{s}\cdot t}\cdot w(\vec{q}_{\alpha},0)
\quad \text{where}\quad B_{\alpha}(t)^{T}  =\mu_{\alpha f}c_{f}\cdot q_{f0}\cdot e^{-\mu_{f}c_{f}\cdot t},
\end{equation}
which denotes ${\alpha}$-th component of the vector $\vec{B}(t)^{T}$.
It clearly shows that this term rapidly diminishes with the time scale of the fast variable, as expected.

Eventually, Eq. (\ref{FullEq}) turns out to be
\begin{equation}
\frac{\partial}{\partial t}w(\vec{q}_{\alpha},t)=\mathcal{L}^{\text{eff}}
\cdot w(\vec{q}_{\alpha},t) + \vec{B}(t)^{T}\cdot \frac{\partial}{\partial\vec{q}_{\alpha}}
e^{\mathcal{L}_{s}\cdot t} w(\vec{q}_{\alpha},0). \label{Feq}
\end{equation}
The question is what the last term in r.h.s. of Eq. (\ref{Feq}), i.e., the inhomogeneous 
term, gives rise to.   As usual, a formal solution is written down,
\begin{equation}
w(\vec{q},t) = \ e^{\mathcal{L}^{\text{eff}}\cdot t}\cdot w^{\text{eff}}(\vec{q},t), 
\end{equation}
with
\begin{equation}
w^{\text{eff}}(q,t )=\left[  1+\int_{0}^{t}dt^{\prime}e^{-\mathcal{L}_{s}\cdot t^{\prime}}\cdot \vec{B}(t^{\prime})^{T}\cdot\frac{\partial}{\partial\vec{q}}\right] w(\vec{q},t=0). \label{weff}
\end{equation}
This means that the reduced system for only slow variables is described by the renormalised 
Smoluchowski operator ${\mathcal{L}^{\text{eff}}}$ with the $t$-dependent ``initial distribution". 
The function  $w^{\text{eff}}(\vec{q},t)$ 
is equal to $w(\vec{q},0)$ at $t=0$, while at $t \gg 1/(\mu_{f}\cdot c_{f})$, i.e., after 
the equilibration of the fast variable,
\begin{equation}
w^{\text{eff}}(\vec{q},t )=  \left[1+\bar{B}\cdot \frac{\partial}
{\partial\vec{q}}\right]w(\vec{q},0)\cong \exp\left[\bar{B}\cdot\frac{\partial}
{\partial\vec{q}}\right]  \cdot w(\vec{q},0),
\end{equation}
where we use again Laplace approximation in the time integration to calculate the 
2nd term in Eq. (\ref{weff}).   $\bar{B}$ denotes a constant shift vector whose 
component is given explicitly,  
\begin{equation}
\bar{B}_{\alpha}  =\frac{\mu_{\alpha f}}{\mu_{f}}q_{f0}, \quad \text{and the initial 
value is effectively shifted as} \quad q_{\alpha 0}^{\text{eff}} 
= q_{\alpha 0} - \bar{B}_{\alpha}.  \label{slip}
\end{equation}
In other words, after the beginning short time, the system evolves as if it has started 
with the {\it slipped initial point.} Now, it should be noted that the slip which 
gives the additional hindrance is given in terms of physical parameters of the system.

Actually, the slip $ - \bar{B}_{R}$ for the radial variable can be estimated for 
the system $^{209}$Bi +$ ^{70}$Zn, for example.    In TCP of LDM, the initial 
point $q_{\it{f} 0}$ of the neck $\varepsilon $ is 1.0 (non-dimensional) 
\cite{Maruhn1972, Sato1979}, and $\frac{\mu_{\alpha f}}{\mu_{f}}$ is estimated with 
the one-body dissipation\cite{Blocki} and at the contact configuration with a 
relaxed neck, say, $\varepsilon $ = 0.2.   The slip in physical dimension is obtained 
with the nuclear radius constant ${r_0}$ = 1.1 fm.    A preliminary value obtained is 
about 2.3fm.   Considering the crude parabolic approximation of LDM, etc, the result 
well explains the phenomenological "injection point parameter {\it s}"( about 2fm) 
in FBD\cite{FBD1,FBD2}.   Analyses of system-dependence etc. will be given elsewhere. 

\subsection{Formation Probability with the initial shift: A New Dynamical Hindrance}

Now we obtain explicit expressions of Formation Probability (Hindrance Factor), 
taking into the initial slip. The solution of Smoluchowski equation with parabolic 
potential and given initial conditions is known as,   
\begin{equation}
w(\vec{q},t) =\frac1{(\sqrt{2\pi})^N} \frac1{\sqrt{\det(\Sigma(t))}} \cdot 
\exp\left[-\frac12 (\vec{q}-e^{-\mu\cdot c\cdot t} \vec{q}_0)^{T}\cdot
\Sigma(t)^{-1}\cdot(\vec{q} - e^{-\mu\cdot c\cdot t} 
\vec{q}_0)\right], 
\end{equation}
with
\begin{equation}
\Sigma(t) =\frac{T}{c}\left(1-e^{-2c\cdot\mu\cdot t}\right)  . 
\end{equation}
Here note that the exponent is $2 \cdot c\cdot\mu$, not $\mu\cdot c$, and that 
the matrix $\Sigma$ is symmetric, so $\Sigma^{-1}$ as well. 

The initial vector $\vec{q}_{0}$ should be the slipped one due to the effective 
initial value given in Eq. (\ref{slip}) and the matrix $\mu$ should be the 
effective one given in Eq. (\ref{mueff}), though it is supposed to be close to 
the original one.

We now consider the system of the two slow variables $(R,\alpha)$ after the 
elimination of the fast neck variable $\varepsilon$. The integration over 
mass-asymmetry $\alpha$ should give a kind of the solution of 1-D Smoluchowski 
equation for the radial coordinate $R$. Then, by integrating over $R$ from minus 
infinity to zero, we obtain the formation probability at time $t$,
\begin{equation}
P_{f}(t)=\int_{-\infty}^{0}dR\text{ }w(R,t)=\frac12\text{erfc}\left[r(t)\right], 
\end{equation}
with the mean trajectory being
\begin{equation} 
r(t)  =\frac{1}{\sqrt{2\Sigma(t)_{RR}}}\exp\left( -\mu\cdot c\cdot t\right) \cdot
\left(\begin{array}[c]{c} R_{0}\\ \alpha_{0} \end{array}\right)_{1}, \label{r}
\end{equation}
where the subscript $ 1 $ denotes the first component of the vector which is the 
product of the exponential matrix and the vector in two dimensions.
The complementary error function is defined as
\begin{equation}
\text{erfc}(x)=\frac2{\sqrt\pi}\int_{x}^{\infty}e^{-t^{2}}dt. 
\end{equation}

The formation probability is explicitly obtained by taking the limit of the time 
${t\rightarrow\infty}$ \cite{Abe2000,Boilley2003}. Here we should remind that 
the origin is taken to be the saddle
point of the LDM potential energy surface. Then all the degrees of freedom 
except the fission valley are confined parabolically. 
Careful evaluations of the time ${t\rightarrow\infty}$ are made on $r(t)$ of 
Eq. (\ref{r}), which gives a simple approximate formula,
\begin{equation}
r(t)\underset{t\rightarrow\infty}{\simeq}\sqrt{\frac{V_{R}}{T}}, \quad \text{with} 
\quad V_{R}=\frac{1}{2}\left\vert c_{11}\right\vert \cdot\bar{R}_{D}^{2} \quad \text{and} \quad \bar{R}_{D}=R_{0}-\frac{\mu_{11}c_{12}+\mu_{12}c_{22}}{\mu_{11}\left\vert c_{11}
\right\vert +\mu_{22}c_{22}}\alpha_{0}. \label{infty}
\end{equation}
Then,
\begin{equation}
P_{\mathrm{form}}\simeq\frac{1}{2\sqrt{\pi}}\sqrt{\frac{T}{V_{R}}}e^{-V_{R}/T}, 
\end{equation}
where the asymptotic expansion of the error function is used for large argument. 
If the second term in r.h.s of Eq. (\ref{infty}) for the expression of $\bar{R}_{D}$   
which depends on $\alpha_{0}$ is negligibly small, the above 
formula is the same as a simple 1-D problem.  It, however, is worth noticing 
here that the initial point above is the slipped one given in Eq. (\ref{slip}), 
i.e., the saddle point height $V_R$ is higher by that like in FBD model.

\section{Concluding Remarks}
We consider the dynamical evolution of di-nucleus systems of very heavy ions as 
relaxation processes of collective degrees of freedom with different time scales, 
starting with the initial conditions given by the incident channel.  
We have analysed effects of the elimination of the fast neck degree on the dynamics 
of the remaining slow variables(mass-asymmetry and radial), starting with N-dimensional 
Smoluchowski equation with the use of N-Z method and of perturbation approximation.     We have 
found that Smoluchowski operator is renormalised and that the inhomogeneous term appears 
to result in a $\it slip$ in the initial values of the slow variables, which provides 
a dynamical origin of the injection point phenomenologically introduced in FBD.   
Since the slip is given in terms of the physical parameters of the incident di-nuclear 
system, we can compare the hindrance effects among various incident channels, 
not only the cold and the hot fusion paths, but also other possible incident 
channels on-going now and in future.

It is noteworthy here that a constant neck value often assumed in numerical 
Langevin calculations is only valid after the relaxation time of the neck.  
Neglect of the dynamics in the beginning short time of the neck relaxation 
should be paid by the slip of the initial values of slow variables.

Instead of N-dimensional Smoluchowski equation, we can also start with K-K 
equation in 2 N-dimensional $(q, p)$ phase space which is equivalent to the 
full Langevin equation used in Ref. \cite{Boilley2019}. For intuition, we 
take two steps, depending on time scales: Firstly, we eliminate the whole 
momentum space, to obtain N-dimensional Smoluchowski equation with memory 
kernel and an inhomogeneous term. Next, we follow the neck elimination 
given in the present article.  We end up with three inhomogeneous terms, 
which, in the lowest order, gives rise to a sum of two shifts from the momentum 
and the neck eliminations, respectively. The results coincide with those obtained 
in \cite{Boilley2019}. They will be published elsewhere, together with systematic 
analysis of incident-channel dependence. The present dynamical approach shall be 
widely developed not only to fusion, but also to quasi-fission etc. It will give 
a novel vista to various dynamical  aspects in heavy-ion collisions as well as to 
quantitative predictions of synthesis cross sections of SHE.

\section*{Acknowledgment}
The authors acknowledge the kind hospitality and the supports given by RCNP, Osaka 
Univ., GANIL, Caen and School of Science, Huzhou Univ., which have enabled us to 
continue our collaborations.  One of the authors (Y.A.) likes to thank Bertrand 
G. Giraud, CEA-Saclay and Bartholom\'e Cauchois, GANIL for useful discussions and 
suggestions in the beginning of the study, and Anthony Marchix, CEA-Saclay for the 
kind invitation to the Nuclear Physics Department which provides opportunity of 
fruitful discussions on SHE production in the future. Part of the work is 
supported by the National Natural Science Foundation of China (NSFC) under 
Grants No. 12075085, 11875125, 11947410, 11790325 and 11790323.

\end{document}